# Anisotropic normal-state properties of the MgB$_2$ superconductor

Pablo de la Mora$^{a,i}$, Miguel Castro$^b$ and Gustavo Tavizon$^b$
$^a$Departamento de Física, Facultad de Ciencias
UNAM, Cd. Universitaria 04510, Coyoacán, D.F., México.
$^b$Departamento de Física y Química Teórica, Facultad de Química
UNAM, Cd. Universitaria, Coyoacán 04510, D.F., México.

**Abstract**

Based on the experimentally-found existence of two gaps in MgB$_2$ (one gap associated to the boron σ-states and the other to the boron π-states), the different contributions to the transport properties, electrical conductivity and Hall coefficient, were studied using the full potential-linearized augmented plane wave method and the generalized gradient approximation. MgB$_2$ doping was analyzed in the rigid band approximation. This permitted the study of the partial substitution of magnesium for aluminium (Mg$_{1-x}$Al$_x$B$_2$) as well as other substitutions such as AB$_2$ (A=Be, Zr, Nb and Ta). The σ bands (boron σ-states), which are associated to the large superconducting gap, are very anisotropic at E$_F$, while the π bands have very little anisotropic character. In (Mg$_{1-x}$Al$_x$B$_2$) T$_c$ diminishes with Al content, the other compounds are not superconductors. In this work it was found that with electron doping, such as Al substitution, the σ-band conductivity decreases and the corresponding bands become less anisotropic. σ-band contribution for BeB$_2$ and ScB$_2$ at E$_F$ is very small and the anisotropy is much lower. For Zr, Nb and Ta there are no σ-bands at E$_F$. These results give a connection between superconductivity and the character of the σ-band; band conductivity and band anisotropy. This gives a plausible explanation for the diminution of T$_c$ with different doping of MgB$_2$.

Keywords: MgB$_2$, superconductivity, electrical conductivity, anisotropy, electronic structure

## 1. Introduction

Soon after superconductivity was discovered in MgB$_2$ the strong anharmonicity of the boron inplane phonons was found. These anharmonic vibrations were given as one of the possible explanations for the high transition temperature in this compound (1). Another possibility was the influence of the two-dimensional character exhibited by the crystal structure on the electronic patterns (2), but electronic structure calculations (3,4) and experimental measurements (5) showed that this material essentially is a bulk conductor with a ratio of 1-10 for the anisotropy in the electrical conductivity in the a/c directions. Soon after, MgB$_2$ produced more surprises; Liu et al. (6) found theoretical evidence that there should be two gaps. One of them is associated to the inplane boron phonons that are strongly coupled to the p$_{x,y}$-orbitals (σ-bands), these σ-orbitals have a high a/c anisotropy, the other gap, associated to the B:p$_z$ and Mg, is three dimensional. The existence of the two gaps has been experimentally confirmed using tunneling spectra (7,8,9,10). In this paper the dimensionality of the different σ and π bands is studied by Density Functional Theory (DFT). The σ bands were found to be highly anisotropic, while the π bands were found to be essentially three-dimensional. Using the rigid band approximation the separate σ and π band-contributions to the electrical conductivity and the Hall coefficient are discussed as function of electron doping in $Mg_{1-x}Al_xB_2$.

---

$^i$ e-mail: delamora@servidor.unam.mx, fax: (+52-55) 5622 4854

## 2. Computational procedure

The electronic structure calculations were done using the *WIEN2k* code (11), which is a Full Potential-Linearized Augmented Plane Wave (*FP-LAPW*) method based on DFT. The Generalized Gradient Approximation of Perdew, Burke and Ernzerhof (12) was used for the treatment of the exchange-correlation interactions. The energy threshold to separate localized and non-localized electronic states was -6 Ry. For the number of plane waves the used criterion was $R_{MT}$ (Muffin Tin radius) × $K_{max}$ (for the plane waves) = 9. The number of k-points used was 19×19×15 (320 in the irreducible wedge of the Brillouin zone). The magnesium muffin-tin radius=$1.8a_0$ and for boron=$1.68a_0$ ($a_0$ is the Bohr radius). For the convergence the charge density criterion was used, with a threshold of $10^{-4}$. A denser mesh of 100×100×76 (34,476 in the irreducible wedge) was used for the evaluation of the electrical conductivity and the Hall coefficients.

## 3. Crystal and electronic structure

The crystal structure of $MgB_2$ is composed by alternating planes of boron and magnesium. The boron planes have a honeycomb arrangement (like graphite but with no displacement) and between two contiguous planes there are the Mg atoms on the line passing through the centre of the boron hexagons, that is, Mg atoms form a hexagonal arrangement. $MgB_2$ has the so-called $AlB_2$ structure (P6/mmm, s. g. 191), with a=3.0864Å, c=3.5215Å.

The band structure at $E_F$ largely reflects the crystal symmetry, which can be easily seen from the Fermi surface structure (figure 1a) and band structure (figure 1b); there are two almost-vertical surfaces around and close to the Γ-A line, their distance is less than 0.31 of the Γ-M distance. The corresponding bands are mainly B:$p_{x,y}$, having almost no magnesium contribution (σ-bands). These bands have little slope in the c direction (Γ-A and L-M) while in the plane (Γ-M-K-Γ and A-L) the slope is large. The corresponding electrical conductivity, as we will prove, should be mainly in the a-b plane, being almost insulating in the c direction.

There are other two Fermi surfaces forming a horizontal honeycomb-like tubular surfaces, with the holes around the Γ-A line, the separation to this line is more than 0.80 the Γ-M distance. One of the surfaces lies in the Γ-M-K plane, surrounding the M-K line; the other in the A-L-H plane and surrounds the L-H line. The corresponding bands are formed by B:$p_z$ with a small magnesium contribution (4) (π-bands), these bands are three-dimensional or bulk-like, with a large slope at $E_F$ in all directions.

Since the Fermi surfaces can be easily separated in k-space the corresponding contribution to the transport properties can also be independently calculated; in the relaxation time approximation the transport properties (conductivity and Hall coefficient) depend on the band structure at $E_F$ only.

*** **Figure 1a.** Fermi surface of $MgB_2$ (13).

*** **Figure 1b.** Band structure of $MgB_2$. Γ-M-K-Γ and A-L correspond to the *a-b* plane directions and Γ-A and L-M to the c direction. The bands with large circles are of σ character, while the ones with dots are of π character.

## 4. Conductivity and Hall coefficient expressions

Within the framework of the relaxation time approximation the transport properties can be estimated from band structure results, using the following expressions (4,14,15):

$$\Delta_{\alpha\beta} = \frac{e^2 \tau(\varepsilon_F, T)}{\Omega_0} \int d^3k \, v_\alpha(k) v_\beta(k) \delta(\varepsilon_F - \varepsilon(k))$$
$$= \frac{e^2 \tau}{\hbar \Omega_0} \int dA_\alpha \sum_i S(v_\alpha^i(k_F)) v_\beta^i(k_F) \tag{1}$$

$$\Delta_{\alpha\alpha} = \frac{e^2 \tau}{\hbar \Omega_0} \int dA_\alpha \sum_i |v_\alpha^i(k_F)| \tag{1a}$$

$$\Delta_{\alpha\beta\gamma} = \frac{e^2 \tau^2}{\hbar \Omega_0} \int d^3k \, v_\alpha(k) [v(k) \times \nabla_k]_\gamma v_\beta(k) \delta(\varepsilon(k) - \varepsilon_F)$$
$$= \frac{e^2 \tau^2}{\hbar^2 \Omega_0} \int dA_\alpha \sum_i S(v_\alpha(k_F)) [v_\alpha(k_F) \frac{dv_\beta}{dk_\beta}\bigg|_{k_F} - v_\beta(k_F) \frac{dv_\beta}{dk_\alpha}\bigg|_{k_F}] \tag{2}$$

where the summation over *i* runs over all the bands at the Fermi energy. Here $\tau$ is the relaxation time, $\Omega_0$ is the normalization volume, $S(v_\alpha(k_F))$ is the sign of $v_\alpha(k_F)$ and $d^3k = dA_\alpha dk_\alpha$, where $A_\alpha$ is the area perpendicular to $k_\alpha$. The second line of Eq. (1) and (2) is obtained from the properties of the delta function, and from them $\Delta_{\alpha\beta}$ and $\Delta_{\alpha\beta\gamma}$ can be easily calculated.

In expressions 1 and 2 the conductivity tensor is $\sigma_{\alpha\beta}$ and the Hall coefficient tensor is (14,15)

$$R_{\alpha\beta\gamma} = \frac{\Delta_{\alpha\beta\gamma}}{\Delta_{\alpha\alpha} \Delta_{\beta\beta}}. \tag{3}$$

From these expressions the conductivity and the Hall coefficient tensor can be calculated, except for τ, from the band structure. Note that τ contains all the temperature dependence. At higher level of approximation additional anisotropy enters from the anisotropy of scattering, but from the cases studied by Allen and co-workers (16,17,18) this turns out to be a surprisingly small effect (a few percent), at least for electron-phonon scattering at $T_c \geq \theta_D$.

What we are interested in this work are in the relative a- and c-direction quantities (σ-band vs. π-band, etc.) and their ratio ($\Delta_{\alpha\alpha}^a/\Delta_{\alpha\alpha}^c$). Therefore the coefficients $e^2\tau/\Omega$ and $e^2\tau^2/\hbar\Omega$ will not be evaluated (will be set to unity) and arbitrary units will be used. The conductivity $\Delta_{\alpha\alpha}$ ($\Delta_{aa}\equiv\Delta_a$, $\Delta_{cc}\equiv\Delta_c$) is related to the plasma frequency ($w_p$) and to the Fermi velocity ($v_F$)(18,19,20);

$$\frac{\Delta_\alpha}{\tau} \sim w_p^2 \sim N(\varepsilon)v_F^2, \qquad (4)$$

where N(ε) is the density of states.

## 5. Results and discussion

One aspect that can be immediately studied is the effect of electron doping. This doping would resemble the effect associated of replacing Mg by Al. That is, since Al has one electron more, additional electrons are added to the system. Mg and Al are almost completely ionized in these compounds (4), this replacement has little effect on the band structure, except for the position of $E_F$. This substitution can be treated within the rigid band approximation, that is, shifting $E_F$ but leaving the band structure unchanged. Satta et al.(3) have done a similar study, but in their treatment they worked with all the bands together, whereas in the present study the different bands are analysed separately. The results of the two studies for the total conductivity and the Hall coefficient are in close agreement (Satta et al. calculated the plasma frequency, which can be related to the conductivity using the Eq 4).

5.1 Electrical conductivity

Regarding the problem of the anisotropic character observed in the band structure of $MgB_2$ and its consequences on the normal state transport properties of this intermetallic compound, there is a simple and intuitive relation arising from the conductivity expression (Eq. 1). Besides the relaxation time approximation, there are two contributions to conductivity in such equation, one of them is the area vector ***dA***, and the other one is the slope of the band, ***v***, at the neighbourhood of the Fermi level. Since both ***dA*** and ***v*** are parallel vectors they should contribute equally to the ratio $v_a/v_c \sim dA_a/dA_c$. From these considerations, after integration:

$$\frac{A_a/A_c}{\sqrt{\Delta_a/\Delta_c}} \sim 1 \qquad (5)$$

where $A_\alpha$ is the Fermi surface area seen from the α direction and $\Delta_\alpha$ is the conductivity in the same direction. This coefficient is equal to unity for a flat Fermi surface; this equality can also be proven for the Fermi surface of parabolic bands ($E= k_x^2/d^2 + k_y^2/e^2 + k_z^2/f^2$).

Equation (5) should be only approximately equal to unity for a general Fermi surface. It gives an intuitive feeling about the anisotropy of the conductivity for a given band, that is, the asymmetry can be estimated by observing the geometry of the Fermi surface or by comparing the slope of the involved bands in different directions.

For example; for the case of the $MgB_2$ bands taken individually this ratio (Eq. 5) is very close to unity; between 0.98 and 1.1, but for all four bands together it deviates more (0.75-1.06) and this is due to the different character of the σ and π bands. For simple Fermi surfaces the expression 5 should be a good estimate.

Now, using this expression to analyze $MgB_2$; the σ Fermi surfaces have a tubular form (figure 1a), therefore seen from above (c-direction) a small area is displayed in a form of a ring, while from the perpendicular direction (a-b plane) a much larger area is span. Therefore a very anisotropic conductivity results from these Fermi surfaces. On the other hand the π surfaces display a large area in all directions and their associated contribution to conductivity should be quite isotropic.

As mentioned above the Fermi surfaces can be easily separated in k space, therefore the conductivity can be calculated for each of the bands independently. The σ bands at $E_F$ are the #7 and #8, (7σ and 8σ). From the band structure plots it can be deduced that 7σ is the one closer to the Γ-A line (see for example Γ-M in figure 1b). The π bands are the #8 and #9 (8π and 9π), the one in the Γ-M- plane is the lower one (8π) (K-M in figure 1a).

From the band structure plot it can be seen that shifting $E_F$ modifies the Fermi surface; going down the σ and π surfaces come closer (see Γ-M and Γ-K in figure 1b) and at ~ -1.2eV they cross and cannot be easily separated any more. On the other hand raising $E_F$ separates σ and π surface, the σ surface disappears at 0.82eV.

The transport properties will be calculated in the [-1.2, 1.5eV] range. The $E_F$ shift is measured in eV, while doping is measured in electrons (e). The eV scale can be translated to electrons using the Density of States plot; the new range now becomes [-0.97, 0.62e]. In the Al full-doped compound, $AlB_2$, the energy shift would be ~2eV (4). The range will not be extended to 2 eV since new three-dimensional bands appear at ~1.7eV. These bands would unnecessarily complicate the analysis since at such high doping range the material is no longer superconducting (21,22,23).

Increasing the doping (figure 1a and 1b) the σ tubular Fermi surfaces shrink towards the Γ-A line until they disappear at the band edge (0.43e), therefore their contribution to the conductivity diminishes (and finally disappears); this is reflected in the conductivity calculations, figure 2a and 2b. From these figures it can also be observed that the 8π-band conductivity also decreases, but it does not disappear, while for 9π it increases. From this perspective, the 7σ, 8σ and 8π bands are hole-like while the 9π-band is electron-like. All the bands have contributions to the conductivity in the *a*-direction of the same order, while in the c-direction the σ-bands contribution is much smaller, about 1/50 of the π-bands contribution.

Our results for total conductivities (σ and π bands together) in the plane and in the c-direction give the same results to those reported by Satta et al. (3).

*** **Figure 2.** Conductivity contribution of different bands: a) a-direction, b) c-direction and c) a/c ratio.

The a/c conductivity ratio is shown in figure 2-c. These results show that the π band anisotropy is small (~1). For the σ bands the anisotropy in quite large; at the highest point (lowest doping) it is of the order of 60. With increasing doping the anisotropy diminishes, at $E_F$ is ≈ 45, and at the band edge it reaches the value of ~ 8. The anisotropy of all the bands taken together (σ and π) at $E_F$ is 1.11, the experimental value is between 3 and 10 (5).

For the high $T_c$ superconductors the calculated a/c ratio in the conductivity is of this order. For example, for $La_{1.85}Sr_{0.15}CuO_4$ it is 27.5 and for $YBa_2Cu_3O_7$ the a/c ratio is 16 and for the b/c ratio is 7 (18). The experimental anisotropy is orders of magnitude larger ($10^2$ to $10^5$). $Bi_2Sr_{2-x}La_xCuO_{6+\delta}$ seems to be most anisotropic material among cuprates = $8·10^5$ (24). Additionally to this anisotropy, the temperature dependence of $\rho_c$ is in most cases semiconducting, $d\rho_c/dT < 0$, whereas that of $\rho_{ab}$ is metallic, $d\rho_c/dT > 0$. The anisotropy in the normal-state conductivity agrees well with a two-dimensional electronic state in the $CuO_2$ plane, while the $\rho_c$ points toward incoherent c-axes transport (25). The much higher experimental value can be explained, at least partially, by the strong correlations of the copper d-orbitals in the copper-oxygen planes that tend to localize the electrons (26).

As it can be seen, the anisotropy in the conductivity is present in all high $T_c$ superconductors, including $MgB_2$. This last material falls in this category due to the fact, as was mentioned above, that superconductivity is sensitive to the band origin of the electrons, and the σ bands are highly anisotropic.

From the results shown in figure 2, it can be seen that by increasing the electron doping the σ-bands contribution diminishes and the anisotropy character of these bands diminish. Therefore the superconductivity in this material should disappear with electron doping, since the main electron-phonon coupling is due to boron inplane phonons (1).

This can be more clearly observed when $MgB_2$ is doped with aluminium (21,22,23); $T_c$ diminishes with the Al content. Putty et al. (23) found a linear relationship between Al content and $T_c$, superconductivity disappearing at *0.58Al*. Our calculations for the σ contribution to conductivity in the a-direction, $\Delta_a^\sigma$, also diminishes linearly, but it vanishes at *0.43e* (figure 2a). This discrepancy could be due to the rigid band approximation used in our methodology.

On the other hand, if this material could be doped with holes, such as in (Na, Mg)B$_2$, the opposite effect would be produced and T$_c$ should increase since $\Delta_a^\sigma$ and the anisotropy increases; de Coss et al. (27) calculated the σ-Fermi surface area and they also predicted a rise of T$_c$ when Na substitutes Mg in MgB$_2$.

Satta et al. also calculated the band structure of BeB$_2$ (3) (an interesting compound to compare with). They found a very similar structure to MgB$_2$, but also important differences: a) E$_F$ is higher, closer to the σ-band edge (lower σ-conductivity); b) the σ band slope in the c-direction is significantly higher, therefore this material should have a less anisotropic σ-conductivity; $\Delta_a/\Delta_c = 7.5$ (for MgB$_2$: $\Delta_a/\Delta_c = 43$). ScB$_2$, on the other hand shows a low T$_c$ (~ 1.5K), this compound is similar to BeB$_2$ in relation to the σ-bands (28). The important difference is that Sc contributes with d-orbitals, this changes the morphology of the bands. A rough estimate, using Eq. 5, gives an anisotropy of ~ 6.5. These differences, in comparison with MgB$_2$, could be important factors that may account for the reduction of T$_c$ in these materials.

ZrB$_2$, NbB$_2$ and TaB$_2$ have the same AlB$_2$ crystalline structure, but their electronic structure is quite different since they have *d* electrons. For instance, there are no anisotropic bands at E$_F$ (4), the *d* electrons add more charge to the compound shifting E$_F$ higher. Even more, the *d* electrons also interact with the B:p$_{x,y}$ orbitals, that is, there are no pure B:σ orbitals anymore, as was the case for MgB$_2$. From the perspective of the present discussion these metallic borides are not expected to be superconductors; this is the case for NbB$_2$ and TaB$_2$. For ZrB$_2$ Gasparov et al. (29) found T$_c$=5.5K (this result has not been confirmed by other experimental papers). The superconductivity of this material may be originated from a different mechanism.

The anisotropy of the B:σ-bands in these AB$_2$ compounds by itself is not responsible for the superconductivity, but it indicates the amount of contamination of the A-element orbitals. This contamination should also perturb the inplane boron phonons, which probably become weaker and less anharmonic. This shows the indirect influence of the anisotropy in the destruction of the superconductivity.

5.2 Hall coefficient

The Hall coefficient, Eq. (3), is non linear in terms of $\Delta_{\alpha\alpha}$ and the Hall coefficient has no direct meaning for the individual bands. Then it is interesting to study the bands-contribution to the term $\Delta_{\alpha\beta\gamma}$. The P6 symmetry in the a-b plane implies that any direction in the plane is equivalent in relation to the transport expressions, while the c direction is different. A good choice of perpendicular coordinates would be (a, y, c), with y in the a-b plane and perpendicular to a. There are two independent terms, $\Delta_{yca}$ (=$\Delta_{cay} \equiv \Sigma_a$) and $\Delta_{ayc}$ ($\equiv \Sigma_c$). The last index in $\Delta_{\alpha\beta\gamma}$ ($\equiv \Sigma_\gamma$) corresponds to the direction of the magnetic field.

The σ-bands contribution to $\Sigma_a$ is small, 1/50 of the π-band contribution (figure 3a). The 7σ, 8σ and 8π are all positive while 9π is negative, that is, the former ones are hole-like and the latter one is electron-like, in agreement with the conductivity results above and those given by Matsui and Tajima (5). On the other hand, figure 3b shows that in $\Sigma_c$ all the contributions are of the same order, but all positive, that is, all are hole-like.

*** **Figure 3.** Contribution to the Hall coefficient ($\Delta_{\alpha\beta\gamma}$) of the different bands: a) the magnetic moment in the plane, b) magnetic moment in c-direction, c) Hall coefficient (the letter refers to the magnetic moment direction).

In figure 3c $R_{yca}$, the hall coefficient, is negative which is mainly due to the large $9\pi$ contribution. The calculated $R_{ayc}/R_{yca} \approx -4.12$ should be compared with the experimental value of the order of 1, reported by Matsui and Tajima (5). It should be mentioned that Satta et al. found the same theoretical results (1).

The discrepancies in the conductivity and the Hall coefficient with the experimental results are probably due to a) $\tau$, the relaxation time, that was taken as constant, while it may have a different value in the a and c directions and this would affect the values of the conductivity and Hall coefficient, b) the use of the rigid band approximation.

## 6. Conclusions

Conductivity measurements show that $MgB_2$ is a fairly isotropic conductor, but the fact that there are two superconducting gaps, one for the σ bands and the other for the π bands, gives a totally different perspective, that is, the electrons seem to distinguish which band they belong to. For this reason the different contributions to the electrical conductivity and Hall coefficients were analyzed. This separation of contributions permitted the study of the individual band-anisotropy and their consequences on the transport properties. The σ Fermi surfaces are very anisotropic, and the associated electrical conductivity is also anisotropic, oppositely the π Fermi surfaces are fairly isotropic.

The σ-bands contribution and their conductivity anisotropy was discussed for several materials in relation to the superconducting properties: a) $ZrB_2$, $NbB_2$ and $TaB_2$ are non superconducting or have a low $T_c$, these materials do not have pure B:σ-band at $E_F$; b) $BeB_2$, also a non-superconductor, has $E_F$ nearer the σ-band edge, also these bands are considerably less anisotropic; c) with Al-doping in $Mg_{1-x}Al_xB_2$ $T_c$ reduces linearly with $x$ and finally disappears at $x=0.58$, $\Delta_a^\sigma$, in our results, follows the same $x$-dependence, but vanishes at *0.43e*, the σ-bands anisotropy also diminishes with electron doping.

The σ-bands contribution and their anisotropy emerge as fundamental factors to the superconducting mechanism of the $MgB_2$ system.

**Acknowledgements**

Support from DGAPA-UNAM under project PAPIIT-IN-101901 is gratefully acknowledged. Also support from María Sabina Ruiz-Chavarría was very valuable.

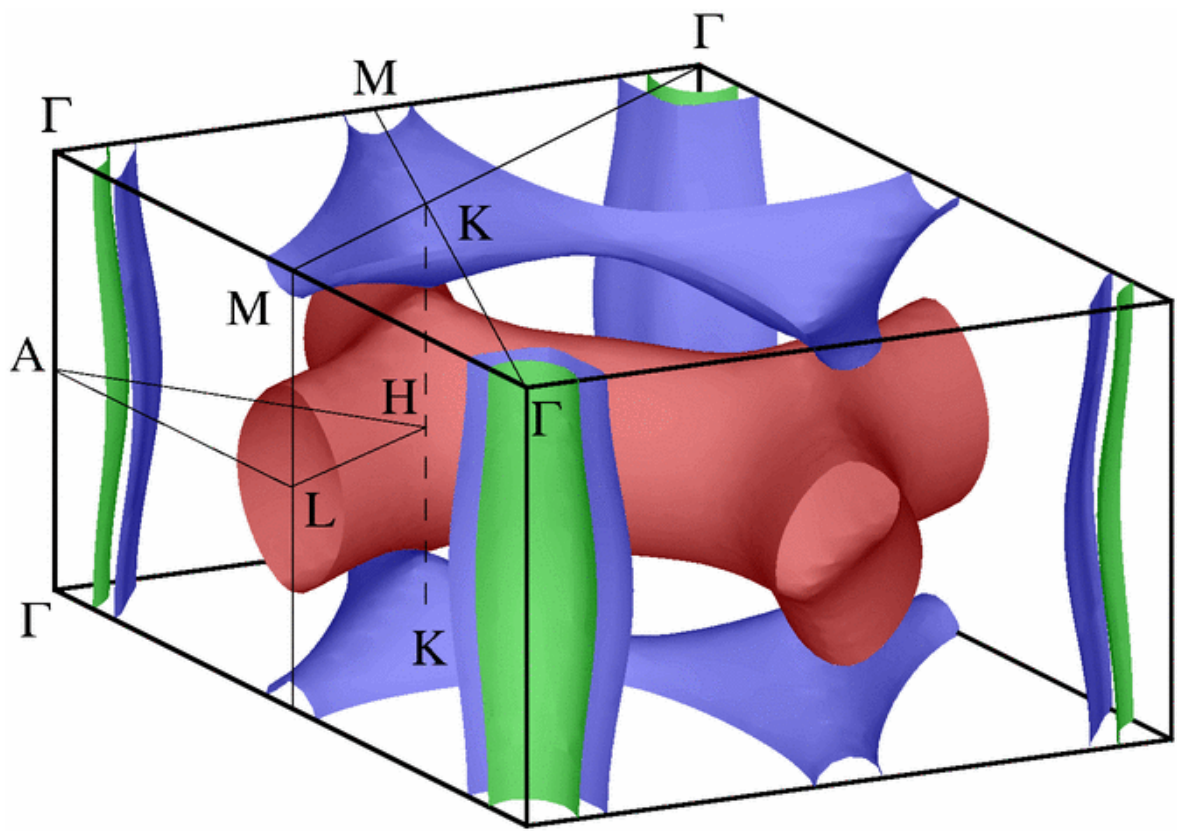

**Figure 1a**

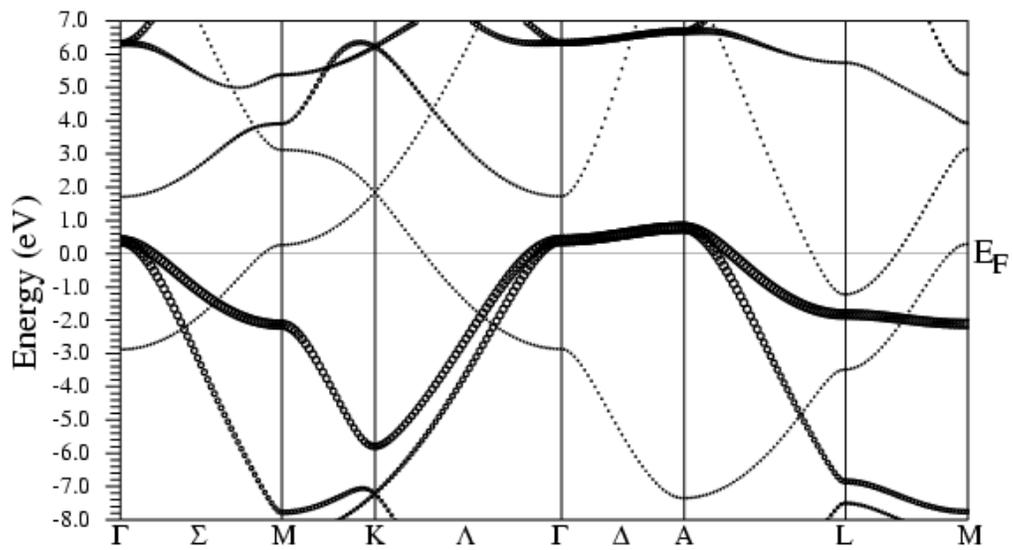

Figure 1b

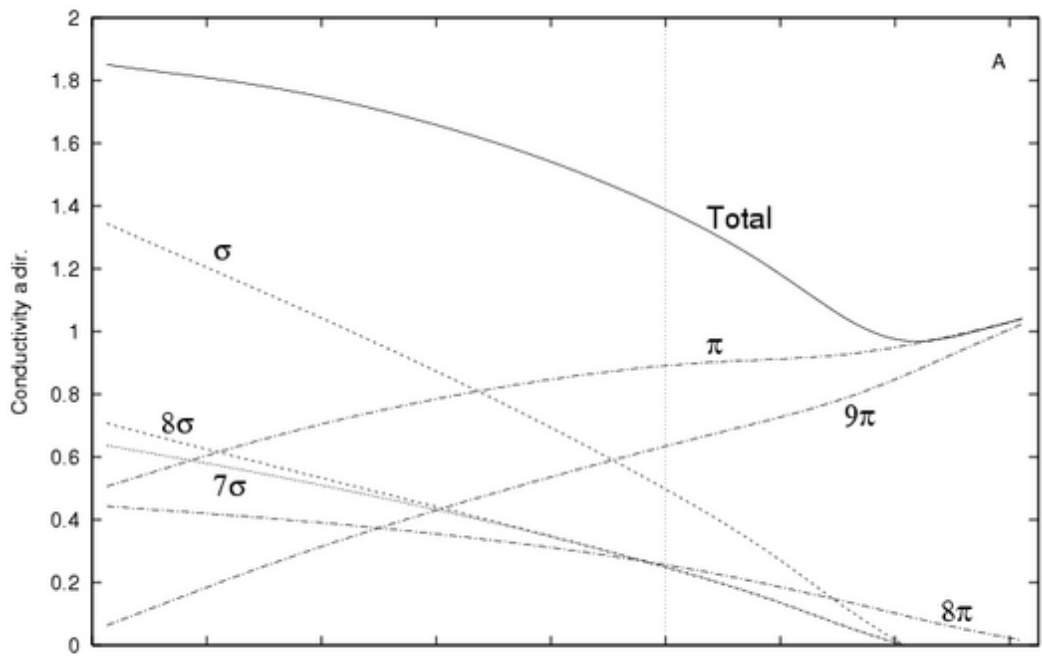

Figure 2a

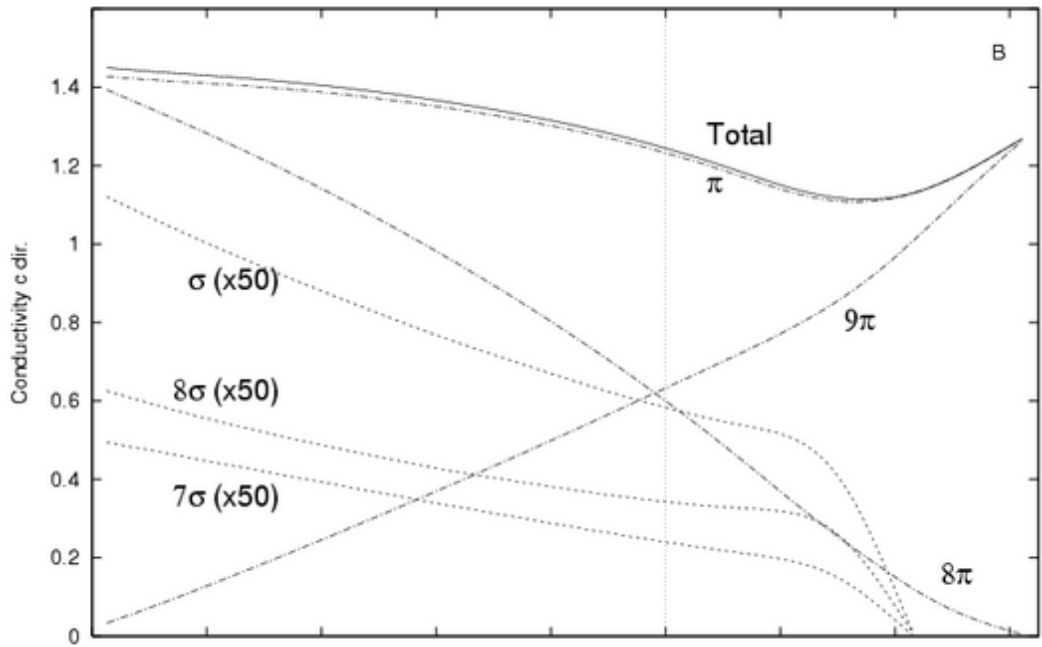

Figure 2b

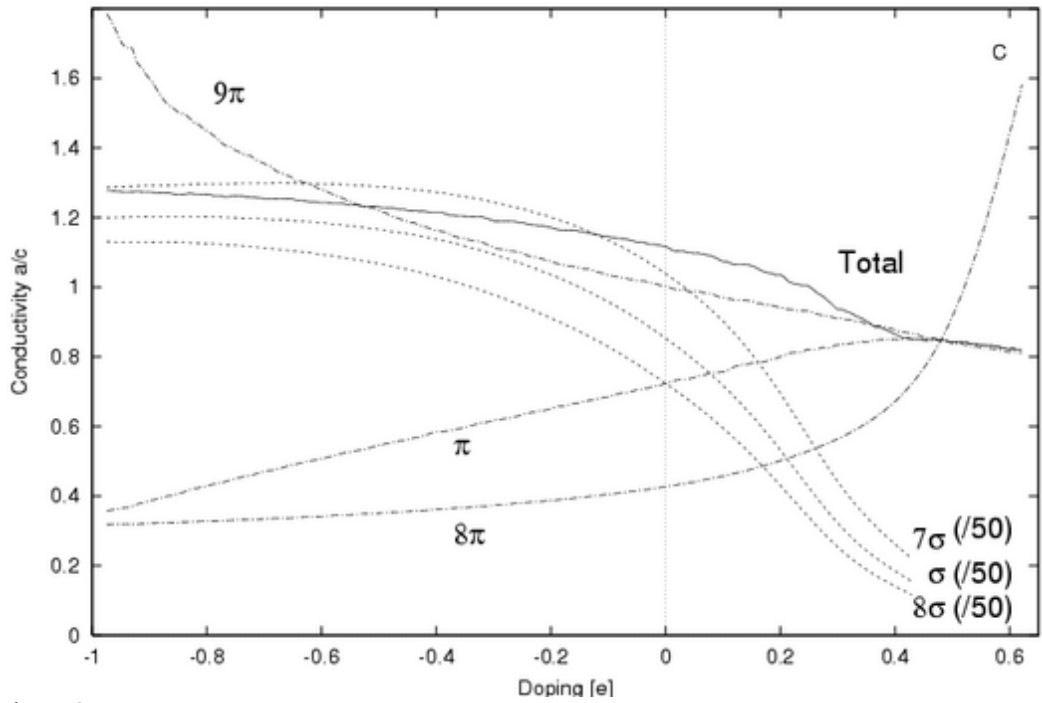

Figure 2c

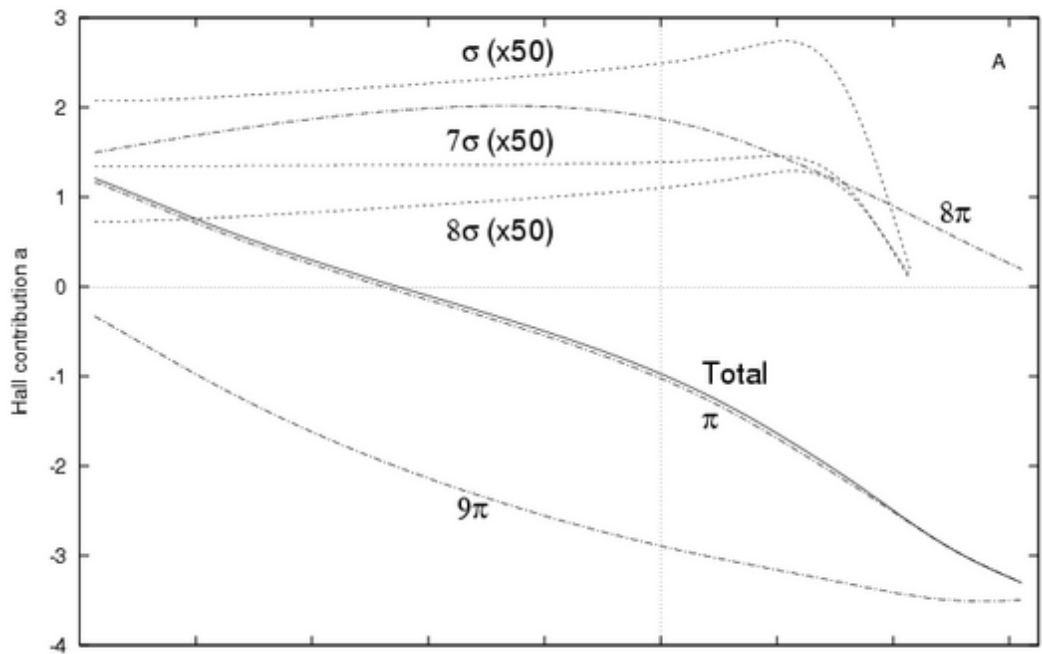

Figure 3a.

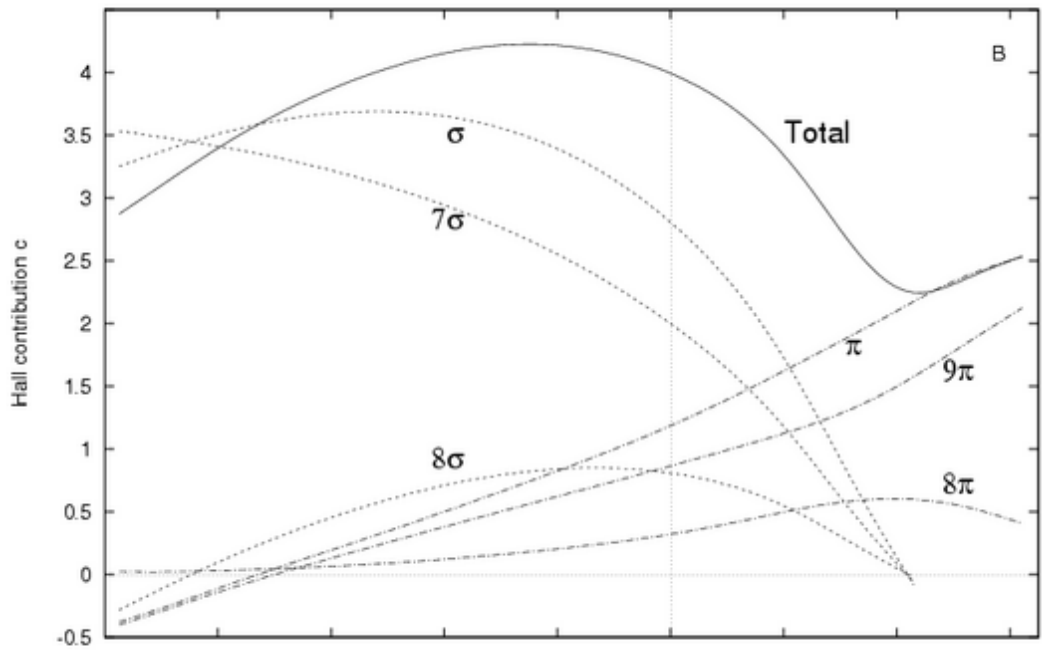

Figure 3b

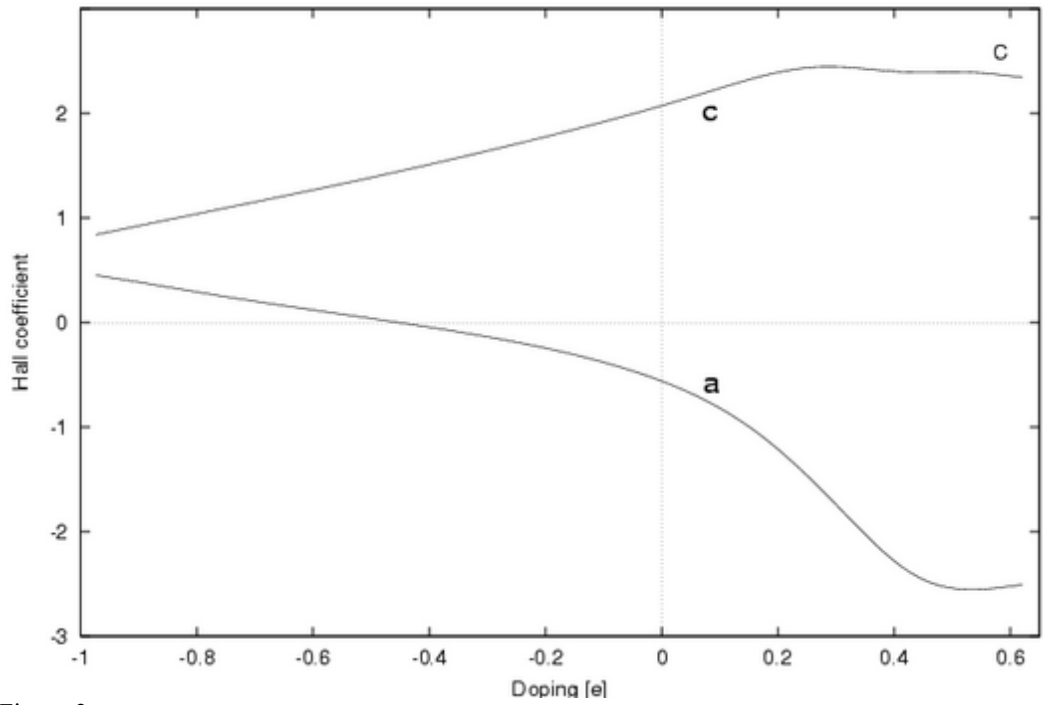

Figure 3c